\documentclass{aa}
\usepackage{graphicx}

\begin{document}

%\thesaurus{10             % ?????????????     
%             (06.07.2;   % Sun: granulation
%              06.16.2;   % Sun: photosphere
%              03.20.1;   % Techniques: image processing 
%              03.20.8)}  % Techniques: spectroscopic

   \title{Non-thermal line-broadening in solar prominences}

   \author{G. Stellmacher\inst{1}
          \and
          E. Wiehr\inst{2}}

   \offprints{E. Wiehr}

   \mail{ewiehr@astrophysik.uni-goettingen.de}

   \institute{Institute d'Astrophysique (IAP), 98 bis Blvd. d'Arago, 
             75014 Paris, France
              \and
              Institut f\" ur Astrophysik der Universit\"at,
              Friedrich-Hund-Platz 1, 37077 G\"ottingen, Germany}

   \date{Received Oct., 2013; accepted ....}

\abstract
{}
{We show that the line broadening in quiescent solar prominences is mainly 
due to non-thermal velocities.}
{We have simultaneously observed a wide range of optically thin lines in 
quiescent prominences, selected for bright and narrow Mg\,b emission without 
line satellites from macro-shifts.}
{We find a ratio of reduced widths, $\Delta\lambda_D/\lambda_0$, of 
H$\gamma$ and H$\delta$ of $1.05\pm0.03$, which can hardly be attributed to 
saturation, since both are optically thin for the prominences observed: 
$\tau_{\gamma} \le0.3$, $\tau_{\delta}\le 0.15$. We confirm the ratio of 
reduced widths of He\,4772\,(triplet) and He\,5015\,(singlet) of $1.1\pm0.05$ 
at higher significance and detect a width ratio of Mg\,b$_2$ and Mg\,4571 
(both from the triplet system) of $1.3\pm0.1$.}
{The discrepant widths of lines from different atoms, and even from the 
same atom, cannot be represented by a unique pair [T$_{kin}$ ; V$_{nth}$]. 
Values of T$_{kin}$ deduced from observed line radiances using models 
indicate low temperatures down to T$_{kin}\approx 5000$\,K. Non-thermal 
velocities, related to different physical states of the respective emitting 
prominence region, seem to be the most important line broadening mechanism.}

\titlerunning{Non-thermal line-broadening in solar prominences}
\maketitle

%
%________________________________________________________________
%

\section{Introduction}

An essential parameter for the description of a prominence plasma is its 
temperature, usually derived from the widths of optically thin lines. 
Widths can be observed with high accuracy and are independent of the 
calibration in absolute units, unlike an analysis using line radiance data. 
The large spectral resolution of ground-based spectrographs (typically 
$\lambda / \Delta \lambda \approx 4\cdot 10^5$) assures a negligible influence 
of the instrumental profile on the narrow prominence emission lines. In 
contrast, existing space-born spectrographs (as e.g. SUMER on-board SOHO), 
designed for broad coronal emissions, strongly affect the widths of prominence 
lines (cf. Stellmacher, Wiehr, Dammasch 2003). 

Prominence emission lines are generally assumed to be broadened by Doppler  
motions resulting from thermal (V$_{th}=\sqrt{2\,R\,T_{kin}/\mu}$) and 
non-thermal (V$_{nth}$) velocities. The separation of both requires (at 
least two) lines from atoms of different mass $\mu$ originating in the 
same prominence volume. Clearly, it is advantageous to use as broad a 
range of lines as possible. However, apart from the first Balmer and 
the Ca\,{\sc ii} lines, prominence emissions in the visible spectrum 
are faint and thus difficult to observe at sufficiently high spectral 
resolution. Extended ground-based observations of faint metallic lines 
at high spectral resolution have been almost exclusively published 
by Landman and collaborators (e.g. Landman 1985). 

Comparing the Doppler widths of Na\,D$_{1,2}$ with those of He\,D$_3$ and 
Ca\,{\sc ii}\,8498, Landman (1981) found that ''the assumption of a common 
emission region for these lines appears to have limited validity''. Recently, 
Ramelli et al. (2012) and Wiehr et al. (2013) resumed such observations and 
found that 'cool' emissions (e.g. of Na\,D or Mg\,b) are accompanied by rather 
broad H$\gamma$ lines of widths corresponding to T$_{kin}>10,000$\,K, and 
conclude that observed Doppler widths cannot simply be described by a unique 
pair [T$_{kin}$; V$_{nth}$]. In Fig.\,1 we summarize the data by Ramelli et al.
(2012; listed in their Table\,3, but here with error bars) as a diagram of
$(c\cdot\Delta\lambda_D/\lambda_0)^2$ versus the inverse atomic mass, showing 
the large deviation from a linear relation expected for a unique [T$_{kin}$; 
V$_{nth}$]. These observations were taken consecutively and might thus 
be influenced by time evolution or drifts of the prominence over the 
spectrograph slit.

%________________________________________________________________
%  Fig.1  [Ramelli 1/My]
   \begin{figure}[h] 
   \hspace{-1mm}
   \includegraphics[width=8.9cm]{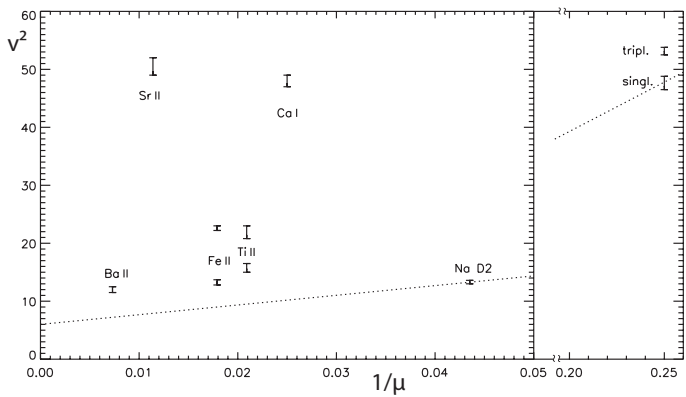}
   \caption{Observed line widths V$^2=(c\cdot\Delta\lambda_D/\lambda_0)^2$ 
           [(km/s)$^2$] versus inverse atomic mass $0\le1/\mu\le0.05$ of 
           metallic lines and  $0.20\le1/\mu\le0.26$ for helium singl. and 
           tripl. (from Ramelli et al. 2012; listed in their Tab.\,3, here with 
           error bars). The dotted line connects Na\,D$_2$ and He\,singlet 
           with apparent $T_{kin}=9950$\,K and $V_{nth}=2.5$\,km/s but 
           does not fit any width of the other emission lines. The range 
           $0.05\le1/\mu\le0.20$ is omitted.}
%\label{Fig1}
    \end{figure}
%________________________________________________________________
%

%________________________________________________________________
%

\section{Observations}
 
\subsection{THEMIS data}

We obtained simultaneous spectra of the prominence emissions H$\gamma$, 
He\,5015\,(singlet) with Fe\,{\sc ii}\,5018, He\,4472\,(triplet) with 
Ti\,{\sc ii}\,4468, He\,{\sc ii}\,4686 and Mg\,b$_1$ (replacing the former 
Mg\,b$_2$ line for technical reason) with five CCD cameras and 10\,sec 
exposure time with the 90\,cm THEMIS telescope on Tenerife in June 2013. 
As for the former observations at the Locarno observatory, IRSOL (cf. 
Ramelli et al. 2012), we selected quiescent prominences with bright but 
narrow Mg\,b emissions without marked macro shifts, thus assuring a 
'spectral photon concentration'. The positions of the observed prominences 
are listed in Table\,1. 

%+++++++++++++++++++++++++++++++
%% Table(1)
\begin{table}[h]
\caption{Date and heliographic position of the prominences.} 
\begin{tabular}{lccc} % l: left, c: center, r: right
\hline 
\hspace{9mm}    & & THEMIS     &                    \\
\hspace{9mm}  1 & June, 06  & 2013 & west limb, $25^o$N \\ 
\hspace{9mm}  2 & June, 07  &  & west limb, $21^o$N \\ 
\hspace{9mm}  3 & June, 10  &  & west limb, $22^o$N \\
\hspace{9mm}  4 & June, 10  &  & west limb, $33^o$S \\
\hspace{9mm}  5 & June, 11  &  & west limb, $22^o$N \\
\hspace{9mm}  6 & June, 11  &  & east limb, $20^o$S \\
\hspace{9mm}  7 & June, 12  &  & west limb, $38^o$N \\ 
\hspace{9mm}  8 & June, 12  &  & west limb, $25^o$N \\ 
\hspace{9mm}  9 & June, 12  &  & west limb, $68^o$S \\
\hspace{9mm}    & &  IRSOL     &                    \\
\hspace{9mm} 10 &  Oct, 09  & 2013 & west limb, $20^o$S \\
\hspace{9mm} 11 &  Oct, 11  &  & west limb, $23^o$N \\
\hspace{9mm} 12 &  Oct, 16  &  & west limb, $05^o$N \\
\hspace{9mm} 13 &  Oct, 16  &  & east limb, $30^o$N \\
\hspace{9mm} 14 &  Oct, 17  &  & west limb, $38^o$S \\
\hspace{9mm} 15 &  June, 30  & 2014 & west limb, $13^o$S \\
\hline  
\end{tabular}
\end{table}
%+++++++++++++++++++++++++++++++

For a sufficiently high signal-to-noise ratio, we chose a slit with of 
1.5\,arcsec corresponding to 1000\,km on the sun. The slit was oriented 
in the direction of atmospheric refraction, i.e. perpendicular to the horizon. 
Wavelength dependent refraction in the earth's atmosphere ($\le0.2$\,arcsec 
through our spectral range 4340\,\AA{}$ < \lambda < 5152$\,\AA{}) then moves 
the prominence structures along the slit, thus preserving them in the different
spectra, where they can readily be considered by suitable spatial shifts of 
the respective scans.

We took spectra of the close (emission free) prominence neighbourhood 
immediately after each exposure for a determination of stray light, mainly 
originating from the telescope mirrors (rather than the earth's atmosphere; 
cf. Stellmacher \& Wiehr, 1970). Additional stray light from the 
spectrograph mirrors was minimized by excluding light from the solar disk. 

For a determination of absolute spectral radiance values, we took spectra of 
the disk centre with an integration time of 1\,sec. Ramelli et. al. (2012) 
describe the reduction procedure. We integrated the line profiles over 15 
pixels in the spatial direction, corresponding to 3.4\,arcsec, i.e. 2500\,km 
on the sun. The final emission line profiles were fitted by Gaussians. 
 
\subsection{Further IRSOL data}

Since THEMIS restricts the combination of lines to $\Delta \lambda <800$\AA{} 
and to appropriate grating orders and since the THEMIS data confirmed the 
former sequential results (see section\,3), we took further spectra at IRSOL 
that allowed us to extend existing observations to more than one line from the 
same atom: Additionally to the lines taken with THEMIS, we observed H$\delta$ 
for comparison with H$\gamma$ (in October 2013) and Mg\,4571 for comparison 
with Mg\,b2 (in June 2014, cf. Table\,1).

\subsection{Accuracy of the line widths}

The errors of the emission line widths arise (a) from the apparent 
continuum level between the emission lines and (b) from the quality of 
the fit. The spurious rest of the stray light (a) depends on the 
correct subtraction of a spectrum from the (emission free) vicinity of the 
prominence. We determine zero as mean (count) level between the emissions. 
For the noisiest profile, Mg\,4571 (see Fig.\,2), the rms of this zero 
level amounts to typically 1\% of the counts in the emission maximum. This 
leads to an error of the Mg\,4571 widths of $\pm7.5$\%. The other emission 
line widths are of much higher accuracy (near 1\%). For the strong Balmer 
emissions the zero-level uncertainty yields a width error well below 1\%. 

In contrast with the accuracy of the zero-level (a), that of the fit (b) is 
one order of magnitude higher, since the optically thin profiles appear to be 
well-defined Gaussians. Figure\,2 gives three spectral scans, each covering 
two neighbouring prominence emissions, showing the high signal-to-noise ratio. 

%________________________________________________________________
%  Fig.2  [Spectra]
   \begin{figure}[h] 
   \hspace{-1mm}
   \includegraphics[width=8.9cm]{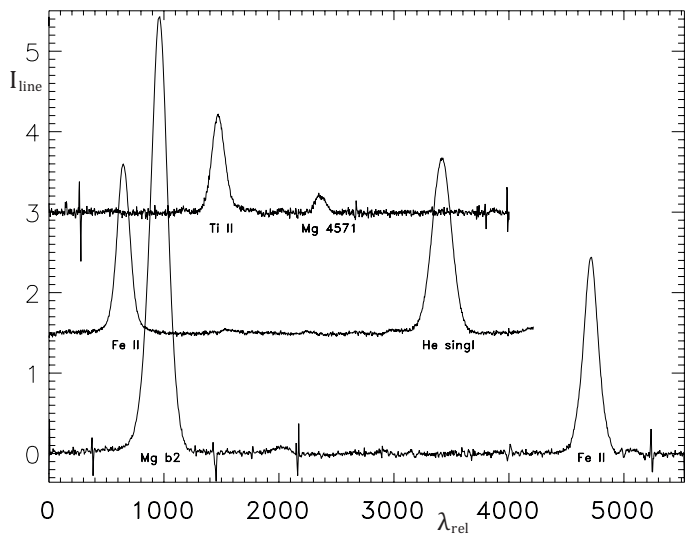} 
   \caption{Spectral line radiance, I$_{line}$[erg/(s\,cm$^2$\,ster\,m\AA{}], 
of Fe\,{\sc ii}\,5168 with Mg\,b$_2$\,5172 (lower), He\,5015 (singl.) 
with Fe\,{\sc ii}\,5018 (middle, ordinate shifted by 
+1.5\,erg/(s\,cm$^2$\,ster\,m\AA{}); and Mg\,4571 with
 Ti\,{\sc ii}\,4572\,\AA{} (upper curve, shifted by 
+3.0\,erg/(s\,cm$^2$\,ster\,m\AA{}), observed in prominence-15 
(cf. Table\,1). The abscissa gives relative wavelengths 
[m\AA].}
%\label{Fig2}
    \end{figure}
%________________________________________________________________

Besides Mg\,4571, the lowest signal-to-noise ratio is found for the 
profile of He\,{\sc ii}\,4686\,\AA{}, which is only measurable in extremely 
bright (and rare) prominences. Figure\,3 shows that even this line profile 
can be fitted to an accuracy of $\le10$\%. 

%________________________________________________________________
%  Fig.3  [HeII 4686]
   \begin{figure}[h] 
   \hspace{3mm}
   \includegraphics[width=7.5cm]{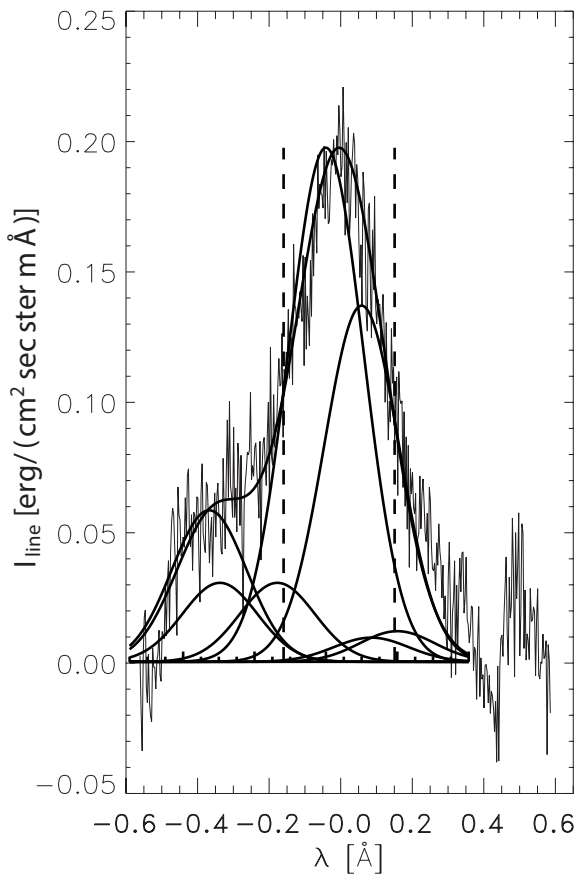} 
   \caption{He\,{\sc ii}\,4686\,\AA{} emission, observed in prominence-15. 
It is fitted by a profile composed of the seven fine-structure components (cf. 
Ramelli et al., 2012) each with FWHM$_{int}=260$\,m\AA{}, yielding a superposed
profile with FWHM$_{tot}=320$\,m\AA{} as observed.}
%\label{Fig3}
    \end{figure}
%________________________________________________________________

\section{Results}

The results from the simultaneous THEMIS observations fully confirm those from
IRSOL in 2011 and 2012, demonstrating that the widths discrepancies found
from sequential data at IRSOL are not an artifact of short-time evolution or
drifts of the prominences over the spectrograph slit. Comparing the observed 
widths of the various emissions from different atoms, we again find with 
THEMIS that none of the sets of lines observed in a given prominence can 
be represented by a unique pair of [T$_{kin} ; V_{nth}$] in the well-known 
relation for Doppler broadening, i.e.

$$(V_D)^2=(c\cdot\Delta\lambda_D/\lambda_0)^2= 2RT_{kin}/\mu+V_{nth}^2$$

\noindent
where $\mu$ is the atomic mass and $V_{nth}$ the non-thermal line broadening 
parameter. We establish the ambiguity of T$_{kin}$ and V$_{nth}$ shown in 
Fig.\,1. This finding is even more impressively seen in Fig.\,4, where we 
attribute to each reduced width, $\Delta \lambda_D / \lambda_0$, a possible 
combination [$T_{kin} ; V_{nth}$], i.e. 

$$V_{nth}=V_D^{obs}-\sqrt{2RT_{kin}/\mu}$$ 
%________________________________________________________________
%  Fig.4       [v_nth ; T_kin Diagramm]
   \begin{figure}[h] 
   \hspace{-14mm}
   \includegraphics[width=10.3cm]{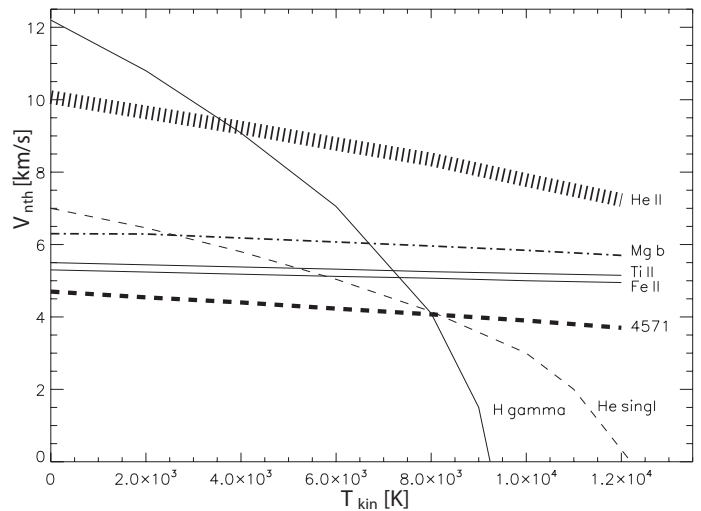}
   \caption{Relation of $V_{nth}$ and $T_{kin}$ for each reduced line width, 
$\Delta \lambda_D^{obs} / \lambda_0$, observed in prominence-15. The thickness of 
each curve represents the error bar of each observed $\Delta \lambda_D / 
\lambda_0$.}
%\label{Fig4}
    \end{figure}
%________________________________________________________________

Figure\,4 shows that line widths of atoms with high mass, $\mu$, are largely
insensitive to T$_{kin}$, in contrast to the Balmer lines. The variety of 
curves excludes a unique $[T_{kin} ; V_{nth}]$ pair fitting the various emission 
line widths. The apparent coincidence of H$\gamma$, He\,singlet and 
Mg\,4571 is rather fortuitous and does not occur in similar diagrams 
for other prominences. Evidently, combinations of lines from atoms of 
different mass, $\mu$, required to separate thermal and non-thermal broadening, 
give contradicting pairs of [$T_{kin}$; $V_{nth}$], which cannot unambiguously 
relate to corresponding common emission regions within the respective 
prominence.

 \subsection{Balmer lines}

The quiescent prominences selected for bright but narrow Mg\,b emission 
persistently show broad H$\gamma$ emissions with reduced widths
$4.1\cdot10^{-5}\le\Delta\lambda_D/\lambda_0 \le 5.2\cdot 10^{-5}$. For purely 
thermal broadening (i.e. $V_{nth} = 0$ km/s), this would correspond to 
$9,200 \le T_{kin} \le 14,800$\,K. 

To get more information about the broadening of Balmer lines, 
we observed H$\delta$ together with H$\gamma$ in Oct. 2013 at IRSOL and 
confirm the large widths of these two optically thin lines. The analysis shows 
that the ratio of the reduced Doppler widths of H$\gamma$ and H$\delta$ is 
larger than 1.0 and does not systematically vary with the H$\gamma$ radiance 
(Fig.\,5).
                         
%________________________________________________________________
%  Fig.5   [Gamma/Delta vs E_tot]
   \begin{figure}[ht]  
   \hspace{-1mm}
   \includegraphics[width=9.0cm]{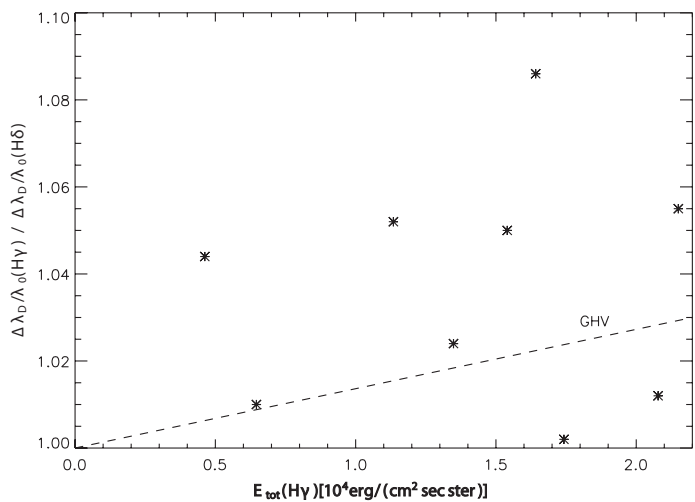}
   \caption{Ratio of observed reduced Doppler widths $\Delta \lambda _D/
   \lambda_0$ of H$\gamma$ and H$\delta$ versus the the total emission ('line 
   radiance') of H$\gamma$. The error bars are smaller than the symbols. 
   Dashes give  a mean of the three models with T$_{kin} =  4300$, 6000 and 
   8000\,K (all V$_{nth}=5$ km/s) by Gouttebroze, Heinzel, Vial (1993).} 
%\label{Fig5}
    \end{figure}
%________________________________________________________________
%

The isothermal model calculations by Gouttebroze, Heinzel, Vial (1993; 
hereafter referred to as GHV) yield a width ratio of H${\gamma}$/H${\delta}$ 
which increases to 1.03 through our observed radiance $E\gamma \le 2.2\cdot 
10^4$ erg/(s\,cm$^2$\,ster). For this range, the models give a total optical 
thickness in the respective emission line centres of $\tau_{\delta}\le 0.15$, 
$\tau_{\gamma} \le0.3$, ($\tau_{\alpha} \le 8$), and predict an increase of 
the mean-width excess H$\gamma$/H$\delta$ with brightness (dashed line in 
Fig.\,5) independent of the assumed model temperature (T$_{kin} = 4300$, 
6000 and 8000\,K; all V$_{nth}=5$ km/s). Our observed width ratios deviate 
from these model calculations and cannot be explained by saturation effects. 

\subsection{Helium lines}
                          
The simultaneous observations with THEMIS confirm those, which have been 
taken consecutively at IRSOL. For the reduced widths 
$\Delta \lambda _D/\lambda_0$, the mean ratio of He\,{\sc ii}\,4686\,\AA{} 
and He\,{\sc i}\,4472\,\AA{} is found to be $1.45\pm0.15$ (former value 1.5)
and that of He\,{\sc i}\,4472\,\AA{} (triplet) and He\,{\sc i}\,5015\,\AA{} 
(singlet) remains $1.1\pm0.05$ (at higher significance). There is no
marked dependence on the brightness of the respective prominence, e.g. on 
the line radiance of He\,{\sc i}\,triplet (Fig.\,6). Thus the variety of 
the width ratios of He\,triplet and He\,singlet cannot be explained by 
saturation effects, (just like the two Balmer lines). Ramelli et al. (2012) 
suggest that the line from the triplet system reflects a hotter origin than 
the singlet line, and is populated via ionization and recombination.

%________________________________________________________________
%  Fig.6        [Tripl/Singl vs E_tot]
   \begin{figure}[h]  
   \hspace{-1mm}
   \includegraphics[width=9.0cm]{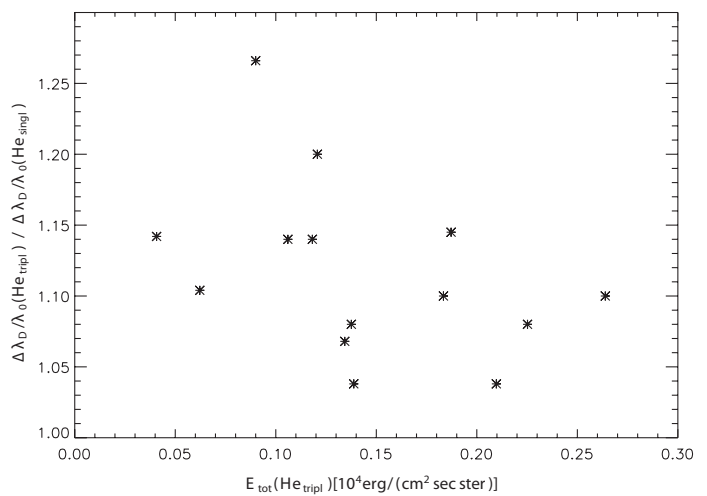}
   \caption{Ratio of observed reduced Doppler widths
    $\Delta \lambda _D/\lambda_0$  of He\,{\sc i}\,4472\,(triplet) 
    and He\,{\sc i}\,5015\,(singlet) versus the total emission ('line 
    radiance') of He\,{\sc i}\,(triplet). The $1\%$ accuracy (cf. 
    Sec.\,2.3) yields error bars of both, the ratio and the triplet radiance 
    values slightly larger than the figure symbols.} 
%\label{Fig6}
    \end{figure}
%________________________________________________________________
%

For the line radiance, we obtain a slightly smaller ratio of 
He\,{\sc i}\,4472 (triplet) and He\,{\sc ii}\,4686 of 45 (formerly 50), 
whereas that of He\,(triplet) and He\,(singlet) of 9.4 (Wiehr et al. 2013) 
remains unchanged.
 
\subsection{Magnesium lines}
 
The faint triplet-to-singlet inter-combination line Mg\,4571\,\AA{}, 
observed by Sotirovski (1985) during eclipse, can only be measured in very 
bright prominences at high spectral resolution. Ramelli et al. (2012) find 
a small reduced width of $\Delta \lambda_D / \lambda_0(4571) =0.9\cdot10^{-5}$. 
We verified this at Locarno in Oct.\,2013 and again find 
$\Delta \lambda _D / \lambda_0 =1.0\cdot10^{-5}$. On June, 30, 2014 we 
observed an exceptionally bright prominence at W/13S ($E_{tot}^{\alpha} 
\approx 4.4\cdot10^5$ erg/(s\,cm$^2$\,ster), i.e. $\tau_{alpha}\approx20$ 
in the GHV tables), where Mg\,4571 had the high line radiance of 
50\,erg/(s\,cm$^2$\,ster). We found that the Mg\,b$_2$ line is 
$1.3\pm0.1$ times broader than Mg\,4571. This cannot be due to saturation, 
since our radiance value of E$_{tot}$(Mg\,b$_2)=1750$\,erg/(s\,cm$^2$\,ster) 
is much smaller than the 6700\,erg/(s\,cm$^2$\,ster) observed by Landman 
(1985), for which he estimates $\tau_0$(Mg\,b$_2)\approx0.7$; besides, our 
Mg\,b$_2$ profile is perfectly Gaussian.   

The width excess of Mg\,b$_2$ over Mg\,4571 cannot be explained in the 
same way as that of He triplet over He singlet, since both upper Mg levels 
belong to the triplet system. The upper level of Mg\,4571 ($3^3P_1$) is 
populated by the Mg\,b$_2$ transition; the large width excess of Mg\,b$_2$ 
over Mg\,4571 is thus hard to understand. On the other hand, the $3^3P_1$ 
level is meta-stable and can be populated from the lowest Mg level ($1^1P_0$; 
singlet) by collisions. In that case, the narrow Mg\,4571 width might indicate 
that these collisions only occur in prominence regions with small line 
broadening.

\section{Discussion}

We have analysed a comprehensive set of faint metallic lines in quiescent 
prominences, which hitherto have rarely been observed at high spectral 
resolution. A sufficiently high signal-to-noise ratio requires prominences 
that show narrow, but bright emissions largely free from macro-shifts (assuring 
a 'spectral photon concentration'). From our former search of measurable 
He\,{\sc ii}\,4686\,\AA{} emission, we know that suitable prominences are 
massive, poorly structured and show pronounced Na\,D emission. The 
simultaneous appearance of 'hot' He\,{\sc ii} and 'cool' Na\,{\sc i} lines
already indicates emissions from different prominence volumes. Our various 
observations with different telescopes, spectrographs and line combinations 
indeed show that the line widths do not allow one to unambiguously separate 
thermal and non-thermal line broadening. A determination of kinetic 
temperatures in quiescent prominences from widths analyses thus remains 
highly uncertain. Given the limits and uncertainties (e.g. Figs.\,1 and 4), 
we are not surprised by the large ranges deduced from widths observed in 
'cool parts of prominences': $6000<T_{kin}<15000$\,K (Jejcic et al., 2014) 
and even $4000<T_{kin}<20000$\,K (Park et al., 2013).

\subsection{T$_{kin}$ from line radiances}

Line radiances offer another way to get information about physical conditions 
in prominences. A remarkably tight relation of H$\alpha$ and H$\beta$ 
emissions over a large brightness range up to E$^{\beta}=1.5\cdot10^5$ 
erg/(s\,cm$^2$\,ster) is found when superposing observations at Tenerife 
with the VTT-spectrograph (Fig.2 in Stellmacher \& Wiehr, 1994), from 2-D 
imaging at the VTT (Fig.\,4 in Stellmacher \& Wiehr, 2000), and from spectra 
with THEMIS (Fig.\,3a in Stellmacher \& Wiehr, 2005). That relation (shown 
in Wiehr \& Stellmacher, 2015) is at best represented by the GHV models with 
T$_{kin}=4300$\,K and T$_{kin}=6000$\,K (both V$_{nth}=5$\,km/s) Gun\'ar et al. 
(2012) use 2D multi-thread models with prominence-corona transition region, 
PCTR, to reproduce observed H$\alpha$ radiances and widths, and deduce 
similar low temperatures down to T$_{kin}\approx 5000$\,K. 

Since none of our various V$_D^2$ versus $1/\mu$ diagrams (e.g. Fig.\,1)
gives gradients corresponding to such low T$_{kin}$, the lines will be 
broadened mainly by non-thermal velocities. Adopting the above 5000\,K, the 
observed H$\gamma$ widths $4.1\le\Delta\lambda_D/\lambda_0\le5.2\cdot10^{-5}$ 
would give $8.3\le V_{nth}\le12.7$\,km/s and the 5\% smaller H$\delta$ widths 
$7.9\le V_{nth}\le12.1$\,km/s; assuming T=5000\,K for the observed Mg lines, 
the Mg\,b$_2$ widths $1.4\le\Delta\lambda_D/\lambda_0\le1.7\cdot10^{-5}$ 
would yield $3.8\le V_{nth}\le4.8$\,km/s, and the 1.3-times narrower Mg\,4571 
line $2.9\le V_{nth}\le3.7$\,km/s.

\subsection{Non-thermal velocities} 

This strong and variable influence of V$_{nth}$ on the prominence emission 
line widths is also supported by the spatial variation of line broadening 
(along the slit; Fig.\,7). If we plot a $1/\mu$ diagram for each spatial 
location (scan-no.), we find nearly parallel curves (Fig.\,8). Although their 
steepness of correspondingly T$_{kin}=10,650$\,K is certainly unrealistic, 
their different ordinate shifts (corresponding to $2\le V_{nth}\le6$\,km/s) 
indicate a spatial variation of non-thermal line broadening. This variation of 
physical states within prominences is also seen from the observed relations 
of He\,D$_3$ and H$\beta$ (Stellmacher \& Wiehr 1995) showing characteristic 
mean ratios for individual prominences which, however, noticeably vary within 
the same prominence (Stellmacher, Wiehr, Hirzberger 2007). 

%________________________________________________________________
%  Fig.7        [Breiten entlang Spalt]
   \begin{figure}[h]  
   \hspace{-1mm}
   \includegraphics[width=9.0cm]{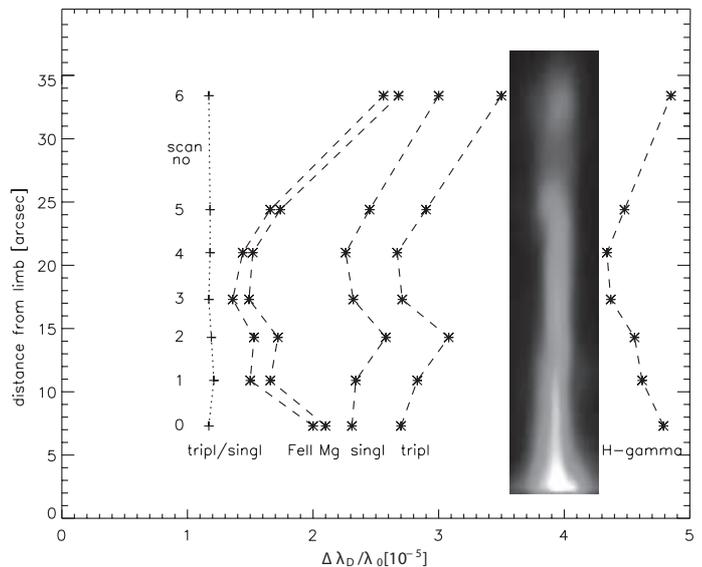}
   \caption{Spatial variation (i.e. along the slit) of reduced Doppler widths 
 $\Delta \lambda _D/\lambda_0$ of H$\gamma$, He\,{\sc i}\,4472\,(triplet), 
 He\,{\sc i}\,5015\,(singlet), Mg\,b$_2$\,5172, and  Fe\,{\sc ii}\,5018; also 
shown the (nearly constant) width ratio of He\,(tripl) and He\,(singl). The 
prominence No.\,1 by chance extends along the direction of refraction, which 
necessarily is the slit orientation, cf. 2.1. Inserted is the Mg\,b$_2$ 
emission spectrum; $\Delta\lambda=0.8$\AA.} 
%\label{Fig7}
    \end{figure}
%________________________________________________________________
%

%________________________________________________________________
%  Fig.8        [1/My entlang Spalt]
   \begin{figure}[h]  
   \hspace{-2mm}
   \includegraphics[width=9.0cm]{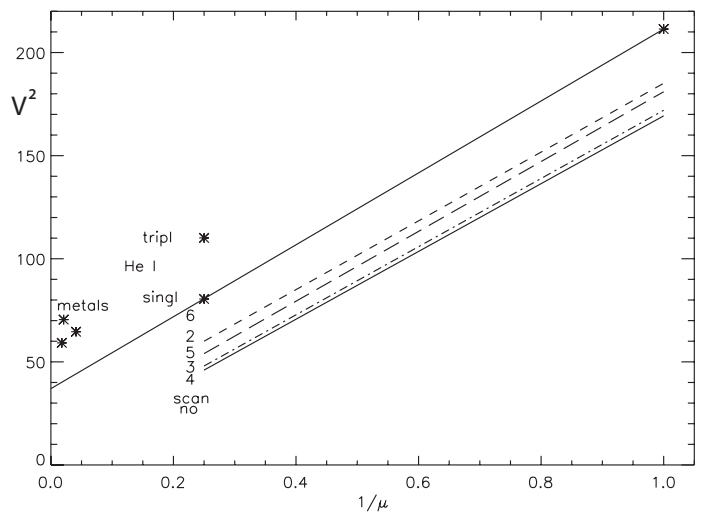}
   \caption{Spatial variation of V$^2=(c\cdot\Delta\lambda_D/\lambda_0)^2$ 
versus inverse atomic mass $1/\mu$ for 5 spatially distinct scans in 
prominence No.\,1. Asterisks give scan-6 (in Fig.\,7); the full line connecting 
H$\gamma$ and He\,{\sc i}\,5015\,(singlet), triplet and metals lay above 
that line (as in Fig.\,1). The other lines similarly connect H$\gamma$ and 
He\,{\sc i}\,5015\,(singlet) values observed for scans-2 through -5 
(in Fig.\,7), and are of nearly the same steepness, T$_{kin}=10,650$\,K; 
their ordinate off-sets correspond to $2\le V_{nth}\le6$\,km/s.} 
%\label{Fig8}
    \end{figure}
%________________________________________________________________
%

A unique pair [T$_{kin} ; V_{nth}$] cannot characterize a mean of 
prominence regions with individual physical state, averaged to assure a 
sufficiently high observational accuracy for the faint (metallic) lines.
In our data we average over 3.4\,arcsec (in the slit direction). If we 
also assume 3.4\,arcsec for the influence of seeing during our (typically 
10\,sec) exposure, the effective spatial area contributing to the line 
profiles is 2500\,km$\times$2500\,km. If we assume that prominence 
structures have widths of $\le150$\,km, separated by spaces of similar 
widths (see e.g. the speckle reconstructed H$\alpha$ image in Wiehr, 
Stellmacher, Hirzberger, 2007; Fig.\,8); our resolution area would 
contain at least 60 of such structures.  

The number of threads along the line of sight may be estimated from 
calculations by GHV: their coolest model with T$_{kin}=4300$\,K does not 
reach our highest emission E$\alpha=4.4\cdot10^5$ erg/(s\,cm$^2$\,ster), 
observed in a massive prominence; their model with T$_{kin}=6000$\,K matches 
this emission assuming D=8000\,km. Adopting this value, one obtains 
$\approx\,27$ threads plus inter-threads (of totally 300\,km) along the 
line of sight, which fairly agrees with Gun\'ar et al. (2012). The total 
prominence volume emitting our observed optically thin lines will then 
represent an integration over $27 \times 60 \approx 1600$ condensations. 
   
Line-specific non-thermal broadening would thus indicate that the condensations 
have different physical states, since a temperature stratification (plus PCTR) 
within clumps of $\le150$\,km, even when adding 27 of them along the line of 
sight, may hardly provide sufficient thickness to emit the observed radiance 
of Na\,{\sc i} through He\,{\sc ii} to 'hot' EUV lines (e.g. Stellmacher, 
Wiehr, Dammasch 2003). Hence, each spectral line originates from a 
sub-entity of the $\approx\,1600$ condensations, which exhibits appropriate 
conditions for its emission and causes a clump-specific non-thermal broadening. 
This is supported by the hierarchy of V$_{nth}$ for H$\gamma$, H$\delta$, 
Mg\,b and Mg\,4571 (see end 4.1). Threads of individual temperature and PCTR 
have already been proposed by Pojoga (1994), which differs from macro-shifted
isothermal threads with equal PCTRs assumed by Gun\'ar et al. (2008).

\subsection{Downdraft motions}

A scenario that predicts plasma clumps at various physical states has 
been modelled by Low et al. (2012). Their calculations show that 
condensations of cool gas must occur in prominences ''as a consequence of 
the non-linear coupling of force balance and energy transport''. The
''catastrophic radiative cooling'' causes the gas to become largely neutral, 
and the frozen-in state breaks down. The clump then sinks and progressively 
''warms up by resistive heating'' (Low et al. 2012) until, at sufficient 
re-ionization, the clump is eventually again frozen-in and stops sinking. 
This ''recurrent break-down of the frozen-in condition'' (Low et al. 2012) 
thus causes a relation between (sink) velocity and temperature. 

If we reasonably assume that some fraction of the vertical falling motion 
shows up as non-thermal velocity (due to turbulence and/or shocks;
cf. Hillier et al. 2012), a sub-entity of the 1600 clumps averaged in our 
spectra, which has the appropriate physical conditions for a specific line 
emission, causes a line-specific non-thermal broadening. 

Downdraft motions in quiescent prominences have been observed with the HINODE 
instrument by Berger et al. (2010) from H$\alpha$ and Ca\,{\sc ii}\,H images. 
These, however, primarily show the spatial variation of local brightenings 
rather than actual motions. We cannot a priori exclude that a change of their 
location reflects a spatial variation of locally optimal excitation conditions. 
Our line widths data, in turn, give strong indication for real plasma 
velocities and can then be considered as an additional support for flows in 
quiescent prominences.

\subsection{Future modelling}

Prominence models may then consider a variety of clumps (of e.g. 150\,km 
diameter) each with individual temperature and non-thermal broadening and 
sum them up along the line of sight (as e.g. done by Gun\'ar et al. 2008) 
and over an area corresponding to the spatial resolution achieved in spectra 
(e.g. our 2500\,km$\times$2500\,km). The observed mean ratios of widths and 
radiance, listed in Table\,2, may be used to scale these models. 

%+++++++++++++++++++++++++++++++
%% Table(2)
\begin{table}[h]
\caption{Mean observed ratios of line broadening, $\Delta \lambda_D / 
\lambda_0$, (upper) and radiance, E$_{tot}$, (lower), valid for 
$0.8\le E_{tot}(H\gamma) \le 5.5\cdot10^4$ erg/(cm$^2$\,s\,ster).} 
\begin{tabular}{lcc} % l: left, c: center, r: right
\hspace{-1mm}  &emission lines & ratios \\
\hline
\hspace{-1mm} &H$\delta$ / H$\gamma$&$1.05\pm0.03$\\ 
\hspace{-1mm} $\Delta\lambda_D/\lambda_0$&He\,{\sc i}\,4472\,(tripl) / 
He\,{\sc i}\,5015\,(singl)&$1.1\pm0.05$ \\ 
\hspace{-1mm} &He\,{\sc ii}\,4686 / He\,{\sc i}\,4472\,(tripl))&$1.45\pm0.3$ \\
\hspace{-1mm} &Mg\,b$_2$\,5172 / Mg\,4571 & $1.3\pm0.3$ \\
\hline
\hspace{-1mm} &He\,{\sc i}\,4472\,(tripl) / He\,{\sc i}\,5015\,(singl)&$9.4\pm2$ \\
\hspace{-1mm} E$_{tot}$&He\,{\sc i}\,4472\,(tripl) / He\,{\sc ii}\,4686 & $45\pm4$  \\
\hspace{-1mm} &Mg\,b$_2$\,5172 / Mg\,4571 & $35\pm4$  \\
\hline 
\end{tabular}
\end{table}
%+++++++++++++++++++++++++++++++

These ratios do not markedly vary through a prominence (here along the 
slit) although the absolute values do (cf. Fig.\,7). We note that the ratios 
are valid for an H$\gamma$ radiance range of $0.8-5.5\cdot10^4$ 
erg/(s\,cm$^2$\,ster), for which the GHV models give $\tau_{alpha} = 2-20$. 
It is still unknown how these ratios will behave in faint prominences with 
pronounced structuring and $\tau_{alpha}<2$. 

%________________________________________________________________
%

\begin{acknowledgements}
We are very indebted to F.\,Hessman for fruitful discussions; C.\,Le\,Men 
contributed valuable support with the THEMIS and M.\,Bianda and R.\,Ramelli 
with the IRSOL observations. We thank an unknown referee for helpful 
suggestions. J.\,Hirzberger (MPS) kindly put his reduction software to our 
disposal. SOLARNET provided financial support. 
\end{acknowledgements}

%________________________________________________________________
%

\end{document}